\documentclass[10pt]{iopart}
\usepackage{iopams}
\usepackage{amsmath_iop}
\usepackage[dvips]{graphicx}
\usepackage{dcolumn}
\usepackage{bm}
\usepackage{CJK}
\usepackage{url}

\def\brakket#1#2#3{\left\langle{#1}\middle|{#2}\middle|{#3}\right\rangle}
\def\ve#1{{\bm{#1}}}

\def\urm#1{\scriptstyle{\text{\textrm{\textmd{\textup{#1}}}}}}

\DeclareMathOperator{\laplace}{\Delta}
 \let\temp\epsilon
\let\temp\epsilon
\let\epsilon\varepsilon
\let\varepsilon\temp
\let\temp\relax

\let\div\relax
\DeclareMathOperator{\div}{div}
\def\eqdef{\mathrel{=:}}
\begin{document}
%
\begin{CJK*}{UTF8}{}
  \title[Relativistic DFT with $ c < \infty $ correction]
  {Relativistic density functional theory with finite-light-speed correction for the Coulomb interaction:
    a non-relativistic-reduction based approach}
  \author{
    Tomoya Naito (\CJKfamily{min}{内藤智也})$ {}^{1, \, 2} $,
    Ryosuke Akashi (\CJKfamily{min}{明石遼介})$ {}^{1} $,
    Haozhao Liang (\CJKfamily{gbsn}{梁豪兆})$ {}^{1, \, 2} $,
    and
    Shinji Tsuneyuki (\CJKfamily{min}{常行真司})$ {}^{1, \, 3} $}
  \address{
    $ {}^{1} $ Department of Physics, Graduate School of Science, The University of Tokyo,
    Tokyo 113-0033, Japan}
  \address{
    $ {}^{2} $ RIKEN Nishina Center, Wako 351-0198, Japan}
  \address{
    $ {}^{3} $ Institute for Solid State Physics, The University of Tokyo,
    Kashiwa 277-8581, Japan}
  \ead{tomoya.naito@phys.s.u-tokyo.ac.jp}
  \date{\today}
  \begin{abstract}
    The Breit correction, the finite-light-speed correction for the Coulomb interaction of the electron-electron interaction in $ O \left( 1/ c^2 \right) $, is introduced to density functional theory (DFT)
    based on the non-relativistic reduction with the local density approximation.
    Using this newly developed relativistic DFT, it is found that the possible outer-most electron of lawrencium atom is the $ p $ orbital instead of the $ d $ orbital,
    which is consistent with the previous calculations based on wave-function theory.
    A possible explanation of the anomalous behavior of its first ionization energy is also given.
    This DFT scheme provides a practical calculation method for the study of properties of super-heavy elements.
  \end{abstract}
  \maketitle
\end{CJK*}
%
\section{Introduction}
\label{sec:intro}
\par
The periodic table of the elements, one of the most fundamental information for general science, is determined by the electronic configuration,
and represents periodicity of fundamental atomic properties, such as the ionization energy and the electron affinity
\cite{
  Mingos2019ThePeriodicTableI_SpringerVerlag}.
Recently, super-heavy elements (SHEs), such as
nihonium ($ Z = 113 $)
\cite{
  Morita2004J.Phys.Soc.Jpn.73_2593},
moscovium ($ Z = 115 $)
\cite{
  Oganessian2004Phys.Rev.C69_021601},
tennessine ($ Z = 117 $)
\cite{
  Oganessian2013Phys.Rev.C87_054621,
  Khuyagbaatar2014Phys.Rev.Lett.112_172501},
and oganesson ($ Z = 118 $)
\cite{
  Oganessian2006Phys.Rev.C74_044602},
were synthesized. 
Syntheses of heavier elements are still ongoing.
However, the positions of the SHEs in the periodic table are yet tentative since their chemical properties have not been established.
\par
A recent experimental measurement of the first ionization energy of lawrencium ($ Z = 103 $)
\cite{
  Sato2015Nature520_209}
casts doubt on the current placement of the SHEs in the periodic table.
The measured value does not follow the tendency of the other $ 5f $-block elements,
which is common to the ionization energy of lutetium ($ Z = 71 $) among the $ 4f $-block elements.
In addition, compared with the vertically neighboring elements in the $ 4f $ and $ 5f $ blocks,
the first ionization energy of lawrencium is smaller than that of lutetium,
whereas those of the other $ 5f $-block elements are larger than those of the corresponding $ 4f $-block elements, respectively.
With these anomalous features of lawrencium---we refer to this as `the puzzle of lawrencium'---it has been discussed in IUPAC
\cite{
  IUPAC}
whether the suitable places of lawrencium and lutetium are the $ f $ or $ d $ block.
In fact, a previous theoretical calculation assuming the electronic configurations referring to the current periodic table did not present these features
\cite{
  Pyykkoe2011Phys.Chem.Chem.Phys.13_161}.
\par
The results mentioned above indicate that the position
of the SHEs on the current periodic table may not generally reflect the actual electronic configurations in their atomic forms and their properties.
Due to short half-lives of the SHEs,
first-principles numerical simulations are essential tools complementary to experiments for
understanding properties of SHEs.
\par
It is known that the relativistic effects are non-negligible in such large $ Z $ atoms
\cite{
  Eides2001Phys.Rep.342_63,
  Schaedel2006Angew.Chem.Int.Ed.45_368,
  Schaedel2014TheChemistryofSuperheavyElements_SpringerVerlag}.
The relativistic effects in electronic systems come from two origins:
(1) difference between the Schr\"{o}dinger and the Dirac Hamiltonians
and
(2) the correction to the Coulomb interaction.
The lowest order of relativistic effects incorporated by using the Dirac equation instead of the Schr\"{o}dinger equation is
$ O \left( \left( Z \alpha \right)^2 \right) \sim O \left( 1/c^2 \right)$.
Here, $ \alpha = e^2 / \left( 4 \pi \epsilon_0 \hbar c \right) \simeq 1/137 $ is the fine-structure constant.
Once the relativistic effect incorporated by the Dirac equation is considered,
the Breit correction between two electrons,
which will be explained later, should also be considered to keep consistency
of the order of $ Z \alpha $.
\par
The relativistic effect due to the correction to the Coulomb interaction can be derived by the quantum electrodynamics (QED) 
through the two-body scattering amplitude
\cite{
  Weinberg1995_CambridgeUniversityPress}.
After the calculation of the quantum field theory in the Coulomb gauge ($ \div \ve{A} = 0 $), 
in the atom with the atomic number $ Z $,
the electron-electron interaction with $ O \left( \left( Z \alpha \right)^2 \right) $ is called the Breit correction
\cite{
  Breit1929Phys.Rev.34_553,
  Breit1930Phys.Rev.36_383},
whereas the Coulomb interaction between two electrons is $ O \left( Z \alpha \right) $.
The Breit correction is also called the relativistic effect or the finite-light-speed effect.
The higher-order terms than the Breit correction are called the QED effects,
which are $ O \left( \left( Z \alpha \right)^n \alpha^m \right) $ with $ n \ge 2 $ and $ m \ge 1 $,
and it is $ \alpha $ times or much smaller than the Breit correction
\cite{
  Eides2001Phys.Rep.342_63}.
Note that the relativistic correction of the Coulomb potential due to the atomic nucleus is
$ O \left( \left( Z \alpha \right)^n \alpha^m \left( m_e / M_{\urm{Nucl}} \right)^l \right) $ with $ n \ge 2 $, $ m \ge 1 $, and $ l \ge 1 $,
where $ m_e $ and $ M_{\urm{Nucl}} $ are the masses of electrons and atomic nuclei, respectively, 
and because of $ m_e / M_{\urm{Nucl}} < 1/1000 $, this effect is negligible
\cite{
  Eides2001Phys.Rep.342_63}.
\par
The first-principles approaches to the electronic properties are classified into two groups:
wave-function theory and
density functional theory (DFT)
\cite{
  Hohenberg1964Phys.Rev.136_B864,
  Kohn1965Phys.Rev.140_A1133,
  Kohn1999Rev.Mod.Phys.71_1253}.
The electronic structure calculation based on 
wave-function theory with the relativistic effects in $ O \left( 1/c^2 \right) $ has been performed
\cite{
  Eliav1995Phys.Rev.A52_291,
  Borschevsky2007Eur.Phys.J.D45_115,
  Sahoo2015Phys.Rev.A91_042507,
  Borschevsky2015Phys.Rev.A91_020501,
  Eliav2015Nucl.Phys.A944_518,
  Pershina2015Nucl.Phys.A944_578,
  Sahoo2016Phys.Rev.A93_032520}.
There, `the puzzle of lawrencium' has also been addressed
\cite{
  Eliav1995Phys.Rev.A52_291,
  Eliav2015Nucl.Phys.A944_518,
  Pershina2015Nucl.Phys.A944_578,
  Desclaux1980J.Phys.France41_943,
  Cao2003J.Chem.Phys.118_487},
which suggested that the outer-most electron of lawrencium is $ p $ electron instead of $ d $ electron.
The wave-function methods, such as the configuration interaction (CI)
\cite{
  Pople1976Int.J.QuantumChem.10_1,
  Pople1977Int.J.QuantumChem.12_149,
  Pople1999Rev.Mod.Phys.71_1267}
and coupled-cluster (CC)
\cite{Coester1958Nucl.Phys.7_421,
  Cizek1966J.Chem.Phys.45_4256,
  Cizek1971Int.J.QuantumChem.5_359}
methods,
are widely used but applicable only to atoms and small molecules due to the large numerical cost,
although they provide, in general, results with high accuracy.
For example,
calculations with the CI or CC with single and double excitations (CISD or CCSD) for $ N $-electron systems
require the numerical cost proportional to $ N^6 $
\cite{
  Jensen2017Introductiontocomputationalchemistry_JohnWiley&Sons,
  Evarestov2012QuantumChemistryofSolids_Springer}.
In contrast, since the numerical cost of DFT is proportional to $ N^3 $ 
\cite{
  Kohn1965Phys.Rev.140_A1133,
  Kohn1999Rev.Mod.Phys.71_1253,
  Jensen2017Introductiontocomputationalchemistry_JohnWiley&Sons,
  Evarestov2012QuantumChemistryofSolids_Springer},
it enables us to simulate larger systems, e.g.,~solids, with moderate accuracy.
Hence, development of DFT with the relativistic effects
with reasonable numerical costs is desired
to study the properties of molecules and coordination complex ions of SHEs, 
whose syntheses are now ongoing under experiment
\cite{
  Nagame2015Nucl.Phys.A944_614,
  Tuerler2015Nucl.Phys.A944_640},
and even solids.
\par
The DFT formalism for QED has been extensively addressed
\cite{
  Rajagopal1973Phys.Rev.B7_1912,
  Rajagopal1978J.Phys.C11_L943,
  MacDonald1979J.Phys.C12_2977,
  Engel2002RelativisticElectronicStructureTheory11_523}.
The exchange--correlation functionals have been proposed with
the \textit{ab initio} local density approximation (LDA)
\cite{
  MacDonald1979J.Phys.C12_2977,
  Kenny1996Phys.Rev.Lett.77_1099}
and the empirical generalized gradient approximation (GGA)
\cite{
  Engel1996Phys.Rev.A53_1367}.
Relativistic extension of the optimized potential method
\cite{
  Engel1998Phys.Rev.A58_964,
  Kreibich1998Phys.Rev.A57_138},
the Dirac DFT with the elimination of the small component
\cite{
  Filatov2002Chem.Phys.Lett.351_259},
and the current-current interaction
\cite{
  Liu1998Phys.Rev.A58_1103,
  Bornemann2012SolidStateCommun.152_85}
have also been established.
\par
In this paper, we take another route to develop a DFT for the relativistic effects.
The non-relativistic reduction of the original Hamiltonian is first performed,
and the Kohn--Sham Hamiltonian is constructed from the reduced Hamiltonian,
instead of the methods adopted in previous works.
This DFT scheme starts from the same Hamiltonian as the wave-function methods for discussion of the electronic structure of super-heavy elements.
Hence, fair comparison with the results of the wave-function method is feasible.
\par
Moreover, the way to construct the DFT scheme is parallel to the conventional non-relativistic (Non-rel) LDA in the PZ81 functional
\cite{
  Perdew1981Phys.Rev.B23_5048}.
The LDA exchange--correlation energy for the reduced Hamiltonian with the relativistic effects up to $ O \left( 1/c^2 \right) $ has been calculated with an accurate numerical solver by Kenny \textit{et al\/.}~\cite{
  Kenny1996Phys.Rev.Lett.77_1099}.
On top of this, we complete the $ O \left( 1/c^2 \right) $ LDA by formulating the Hartree term for the Breit correction to the electron-electron interaction.
Thanks to this formulation,
the relativistic correction is given as additional terms on top of the Non-rel DFT.
Hence,
one can compare the results of the relativistic scheme with those of the Non-rel scheme clearly.
\par
It is known that the LDA, even in the Non-rel DFT, can give erroneous results already for the lighter elements including $ 3d $-electron systems
\cite{
  Vosko1989Phys.Rev.A39_446,
  Lagowski1989Phys.Rev.A39_4972,
  Vukajlovic1991Phys.Rev.B43_3994,
  Martin2004_CambridgeUniversityPress,
  Kosugi2018J.Chem.Phys.148_224103},
which is admittedly cured by the gradient
\cite{
  Engel2002RelativisticElectronicStructureTheory11_523,
  Perdew1996Phys.Rev.Lett.77_3865,
  Engel2011_Springer-Verlag},
the discontinuity of the exchange--correlation potential $ V_{\urm{xc}} $
\cite{
  Engel2011_Springer-Verlag,
  Perdew1982Phys.Rev.Lett.49_1691,
  Engel1992Z.Phys.D23_7,
  Vydrov2007J.Chem.Phys.126_154109,
  Fiolhais2003APrimerinDensityFunctionalTheory_Springer-Verlag}
and self-interaction corrections
\cite{
  Perdew1981Phys.Rev.B23_5048,
  Perdew1979Chem.Phys.Lett.64_127,
  Zhang1998J.Chem.Phys.109_2604}.
Nevertheless,
we expect that our LDA-based scheme is efficient for heavy elements,
since the correction to the exchange energy by those become irrelevant in the large $ Z $ regime
\cite{
  Perdew2006Phys.Rev.Lett.97_223002},
and it is still meaningful to use atomic systems as benchmark calculations.
\par
This paper is organized as follows: 
first, in section \ref{sec:theor}, the theoretical framework of DFT with the Breit correction is introduced.
Then, in sections \ref{subsec:calc_1} and \ref{subsec:calc_2},
all-electron calculation of selected atoms are performed as a benchmark.
In section \ref{subsec:calc_Lr}, the possible reason for `the puzzle of lawrencium' is suggested,
with performing all-electron calculation of lutetium and lawrencium atoms.
Finally, in section \ref{sec:conc}, the conclusion and perspectives of this paper are shown.
In \ref{sec:app_xc}, the detailed discussion about the relativistic exchange--correlation functional is shown, and in \ref{sec:app_hartree},
derivation of the relativistic Hartree term is shown.
%
%
\section{Theoretical Framework}
\label{sec:theor}
\par
In this section, the relativistic DFT with the Breit correction is formulated.
We start from the Dirac equation instead of the Schr\"{o}dinger equation,
where the Breit correction is considered in the electron-electron interaction $ V_{\urm{int}} $.
To use the Kohn--Sham scheme, the Hartree term $ E_{\urm{H}} $ and the exchange--correlation functional $ E_{\urm{xc}} $ should be reconstructed,
since $ V_{\urm{int}} $ is no longer the original Coulomb interaction.
\par
In this paper, the Hartree atomic unit is used,
i.e., $ m_e = \hbar = 4 \pi \epsilon_0 = e^2 = 1 $ and $ c = 1 / \alpha $,
and the Coulomb-Breit interaction refers the electron-electron interaction with the Breit correction as well as the Coulomb interaction.
\subsection{Original Hamiltonian}
\par
In general, the Dirac Hamiltonian in quantum many-body problems reads
\begin{equation}
  \label{eq:Hamil}
  \hat{H}
  =
  \hat{T}
  +
  \sum_j
  V_{\urm{ext}} \left( \ve{r}_j \right)
  +
  \sum_{j < k}
  V_{\urm{int}} \left( \ve{r}_j, \ve{r}_k \right),
\end{equation}
where $ \hat{T} $ is the kinetic operator, 
$ V_{\urm{ext}} $ is the external potential, 
and $ V_{\urm{int}} $ is the interaction between electrons.
Note that 
the Hamiltonian \eqref{eq:Hamil} operates to the Dirac spinor 
and $ \hat{T} $ is the Dirac kinetic operator $ \hat{T}^{\urm{D}} $
instead of the Schr\"{o}dinger kinetic operator $ \hat{T}^{\urm{S}} $.
The Dirac kinetic operator $ \hat{T}^{\urm{D}} $ is written in sum of the single-particle Dirac kinetic operator for electrons $ \hat{t}_j^{\urm{D}} $:
\begin{equation}
  \label{eq:kin}
  \hat{T}^{\urm{D}}
  =
  \sum_j
  \hat{t}_j^{\urm{D}} ,
\end{equation}
where 
\begin{equation}
  \label{eq:Dirac}
  \hat{t}_j^{\urm{D}}
  =
  \beta_j c^2 + c \ve{\alpha}_j \cdot \ve{p}_j.
\end{equation}
Here, $ \ve{\alpha}_j $ and $ \beta_j $ are the Dirac matrices for the $ j $th electron,
\begin{equation}
  \label{eq:alpha}
  \ve{\alpha}
  =
  \left(
    \begin{pmatrix}
      O_2 & \sigma_x \\
      \sigma_x & O_2
    \end{pmatrix}, 
    \begin{pmatrix}
      O_2 & \sigma_y \\
      \sigma_y & O_2
    \end{pmatrix}, 
    \begin{pmatrix}
      O_2 & \sigma_z \\
      \sigma_z & O_2
    \end{pmatrix}
  \right),
  \qquad
  \beta 
  =
  \begin{pmatrix}
    I_2 & 0 \\
    0 & -I_2
  \end{pmatrix},
\end{equation}
where $ \sigma_x $, $ \sigma_y $, and $ \sigma_z $ are the Pauli matrices,
and 
$ O_2 $ and $ I_2 $ are the $ 2 \times 2 $ zero and identity matrices, respectively.
\par
The Coulomb-Breit interaction
\begin{equation}
  \label{eq:int}
  V_{\urm{int}} \left( \ve{r}_j, \ve{r}_k \right)
  =
  \frac{1}{r_{jk}}
  -
  \left[
    \frac{c \ve{\alpha}_j \cdot c \ve{\alpha}_k}{2c^2 r_{jk}}
    +
    \frac{
      \left(c \ve{\alpha}_j \cdot \ve{r}_{jk} \right)
      \left(c \ve{\alpha}_k \cdot \ve{r}_{jk} \right)}{2c^2 r_{jk}^3}
  \right]
\end{equation}
is adopted for the electron-electron interaction $ V_{\urm{int}} $,
where $ \ve{r}_{jk} = \ve{r}_j - \ve{r}_k $, $ r_{jk} = \left| \ve{r}_{jk} \right| $.
The first term is the original Coulomb interaction
and the second term is the Breit correction
\cite{
  Breit1929Phys.Rev.34_553,
  Breit1930Phys.Rev.36_383}.
The kinetic operator for the nuclei is neglected,
and $ V_{\urm{ext}} $ is the interaction between the atomic nucleus and electrons.
Only the Coulomb interaction is considered for $ V_{\urm{ext}} $,
since the finite-light-speed correction to $ V_{\urm{ext}} $ is proportional to $ m_e / M_{\urm{nucl}} $ much smaller than that to $ V_{\urm{int}} $
\cite{
  Eides2001Phys.Rep.342_63}:
\begin{equation}
  \label{eq:ext}
  V_{\urm{ext}} \left( \ve{r}_j \right) = - \frac{Z}{r_j}.
\end{equation}
\subsection{Non-Relativistic Reduction}
\par
As mentioned in section \ref{sec:intro},
the Kohn--Sham Hamiltonian is constructed after the non-relativistic reduction of the original Dirac Hamiltonian \eqref{eq:Hamil},
instead of the method used in the previous works.
\par
According to the Hohenberg-Kohn theorem
\cite{
  Hohenberg1964Phys.Rev.136_B864},
the universal functional $ F $ of the electron density $ \rho $
with respect to the kinetic operator $ \hat{T} $ and the interaction $ V_{\urm{int}} $
gives the ground-state energy of the Hamiltonian via
\begin{equation}
  \label{eq:Egs}
  E \left[ \rho \right]
  =
  F \left[ \rho \right]
  +
  \int
  \rho \left( \ve{r} \right) \,
  V_{\urm{ext}} \left( \ve{r} \right) \,
  d \ve{r}.
\end{equation}
The exchange--correlation energy functional $ E_{\urm{xc}} $ is defined with this $ F $, as mentioned later.
The standard functionals,
such as the PZ81
\cite{
  Perdew1981Phys.Rev.B23_5048}
and PBE
\cite{
  Perdew1996Phys.Rev.Lett.77_3865}
functionals,
are applicable only to the Schr\"{o}dinger scheme.
In the present case, the exchange--correlation functional should be reconstructed on the basis of the Dirac Hamiltonian.
\par
Since only positive-energy states are usually interested, non-relativistic reduction of the Hamiltonian is used for this scheme.
One of the most widely used non-relativistic reduction methods is the Foldy-Wouthuysen transformation
\cite{
  Pryce1948Proc.RoyalSoc.Lond.A195_62,
  Foldy1950Phys.Rev.78_29,
  Tani1951Prog.Theor.Phys.6_267,
  Foldy1952Phys.Rev.87_688}.
The Foldy-Wouthuysen transformation of the Hamiltonian given in equation \eqref{eq:Hamil} derived by Kenny \textit{et al\/.}~\cite{
  Kenny1995Phys.Rev.A51_1898}
is
\begin{align}
  \hat{H}_{\urm{FW}}
  = & \, 
      \hat{T}^{\urm{S}}
      +
      \sum_j
      V_{\urm{ext}} \left( \ve{r}_j \right)
      + 
      \sum_j
      \left\{
      V'_1 \left( \ve{r}_j, \ve{s}_j \right)
      -
      \frac{Z}{2c^2} \frac{1}{r_j^3}
      \ve{s}_j \cdot \left[\ve{r}_j \times i \nabla_j \right]
      \right\}
      \notag \\
    & + 
      \sum_{j < k}
      \frac{1}{r_{jk}}
      + 
      V'_2 \left( \ve{r}_j, \ve{r}_k, \ve{s}_j, \ve{s}_k \right),
\end{align}
where the correction terms $ V'_1 $ and $ V'_2 $ read
\begin{align}
  & V'_1 \left( \ve{r}_j, \ve{s}_j \right)
    \notag \\
  = & 
      - \frac{\nabla_j^4}{8c^2}
      + \frac{Z \pi}{2c^2} \delta \left( \ve{r}_j \right),
      \label{eq:V1} \\
  & V'_2 \left( \ve{r}_j, \ve{r}_k, \ve{s}_j, \ve{s}_k \right)
    \notag \\
  = & 
      - \sum_{j < k}
      \frac{\pi}{c^2} \delta \left( \ve{r}_j - \ve{r}_k \right)
      - \sum_{j < k}
      \frac{1}{2c^2}
      \overleftarrow{\nabla}_j \cdot
      \left[
      \frac{1}{r_{jk}}
      +
      \frac{\left( \ve{r}_j - \ve{r}_k \right)\left( \ve{r}_j - \ve{r}_k \right)}
      {r_{jk}^3}
      \right]
      \cdot \overrightarrow{\nabla}_k \notag \\
  &
    - \sum_{j < k}
    \frac{8 \pi}{3c^2}
    \delta \left( \ve{r}_j - \ve{r}_k \right) \,
    \ve{s}_j \cdot \ve{s}_k 
    - \sum_{j < k}
    \frac{1}{c^2}
    \ve{s}_j \cdot
    \left[
    \frac{3\left( \ve{r}_j - \ve{r}_k \right)\left( \ve{r}_j - \ve{r}_k \right)}
    {r_{jk}^5}
    -
    \frac{1}{r_{jk}^3}
    \right]
    \cdot \ve{s}_k  \notag \\
  &
    + \sum_{j \ne k}
    \frac{1}{c^2}
    \frac{1}{r_{jk}^3}
    \ve{s}_j \cdot
    \left[
    \left( \ve{r}_k - \ve{r}_j \right)
    \times i \nabla_k
    \right]
    + \sum_{j \ne k}
    \frac{1}{c^2}
    \frac{1}{r_{jk}^3}
    \ve{s}_k \cdot
    \left[
    \left( \ve{r}_k - \ve{r}_j \right)
    \times i \nabla_k
    \right].
    \label{eq:V2}
\end{align}
The first and second terms of $ V'_2 $ correspond to the electron-electron Darwin and retardation terms, respectively.
As long as spin-unpolarized systems, such as the closed-shell atoms,
are concerned, the second and third lines of equation \eqref{eq:V2} vanishes.
\par
From the point of view of the Hohenberg-Kohn theorem,
the functional $ F $ is universal
within the Schr\"{o}dinger scheme for the interaction
$ 1/r_{ij} + V'_2 \left( \ve{r}_j, \ve{r}_k, \ve{s}_j, \ve{s}_k \right) $.
\subsection{DFT with finite-light-speed correction}
\label{subsec:finite}
\par
In order to reformulate DFT on the basis of $ \hat{H}_{\urm{FW}} $, 
the Hartree term $ E_{\urm{H}} $ and the exchange--correlation functional $ E_{\urm{xc}} $ in this scheme is derived.
Here, the universal functional $ F $ in equation \eqref{eq:Egs} is separable into four parts
\begin{equation}
  F \left[ \rho \right]
  =
  T_0 \left[ \rho \right]
  +
  \frac{1}{2} 
  \iint 
  \frac{\rho \left( \ve{r} \right) \, \rho \left( \ve{r}' \right)}{\left| \ve{r} - \ve{r}' \right|}
  \, d \ve{r} \, d \ve{r}'
  +
  E_{\urm{Hrel}} \left[ \rho \right]
  + 
  E_{\urm{xc}} \left[ \rho \right],
\end{equation}
where $ T_0 $ is the kinetic energy for non-interacting systems, 
the second term is the Hartree term with the Coulomb interaction,
the third term is the relativistic correction for the Hartree term,
and the fourth term is the exchange--correlation term, which includes the effects of $ V'_2 $ as well as the Coulomb interaction.
\par
The LDA exchange--correlation functional $ E_{\urm{xc}} $ for this interaction
$ \sum_{j < k} 1/r_{jk} + V'_2 \left( \ve{r}_j, \ve{r}_k, \ve{s}_j, \ve{s}_k \right) $ 
has been derived by Kenny \textit{et al\/.}~\cite{
  Kenny1996Phys.Rev.Lett.77_1099}:
this exchange--correlation energy density $ \epsilon_{\urm{xc}} $ is written as
\begin{align}
  \label{eq:Kenny}
  \epsilon_{\urm{xc}}
  \left( r_{\urm{s}} \right)
  & = 
    \epsilon_{\urm{xc}}^{\urm{Non-rel}}
    \left( r_{\urm{s}} \right)
    +
    \frac{9}{8c^2 r_{\urm{s}}^3}
    f \left( r_{\urm{s}} \right) , \\
  f \left( r_{\urm{s}} \right)
  & =
    \begin{cases}
      0.9918
      - 0.29020 r_{\urm{s}}
      + 0.14474 r_{\urm{s}}^2
      - 0.02573 r_{\urm{s}}^3
      + 0.001634 r_{\urm{s}}^4
      &
      \text{($ r_{\urm{s}} \le 5 $)}, \\
      0.75 + 0.044 r_{\urm{s}}
      &
      \text{($ r_{\urm{s}} > 5 $)},
    \end{cases}
\end{align}
where the exchange--correlation energy density $ \epsilon_{\urm{xc}} $ satisfies
\begin{equation}
  E_{\urm{xc}} \left[ \rho \right]
  = 
  \int 
  \epsilon_{\urm{xc}} \left( r_{\urm{s}} \left( \ve{r} \right) \right) \,
  \rho \left( \ve{r} \right) 
  \, d \ve{r},
\end{equation}
and $ r_{\urm{s}} $ is the Wigner-Seitz radius defined by 
\begin{equation}
  r_{\urm{s}} 
  =
  \left( 
    \frac{3}{4 \pi \rho}
  \right)^{1/3}.
\end{equation}
Here, $ \epsilon_{\urm{xc}}^{\urm{Non-rel}} $ is the LDA exchange--correlation energy density in the Non-rel scheme,
and in this calculation the PZ81 functional
\cite{
  Perdew1981Phys.Rev.B23_5048}
is used.
\par
Relativistic correction of the Hartree term, $ E_{\urm{Hrel}} $, is constructed from the first line of equation \eqref{eq:V2}.
The first and second terms of equation \eqref{eq:V2} represent the Darwin term and retardation effect, respectively.
Relativistic correction to the Hartree energy and potential corresponding to the first term of equation \eqref{eq:V2} are
\begin{equation}
  \label{eq:EH1}
  E_{\urm{Hrel}} \left[ \rho \right]
  =
  - \frac{\pi}{2c^2}
  \int
  \left[
    \rho \left( \ve{r} \right)
  \right]^2 \, d \ve{r},
  \qquad
  V_{\urm{Hrel}} \left( \ve{r} \right)
  =
  - \frac{\pi}{c^2}
  \rho \left( \ve{r} \right),
\end{equation}
respectively.
In contrast, the contribution of the second term to $ E_{\urm{Hrel}} $ is proved to be zero
(see \ref{sec:app_hartree}).
The physical meaning of this vanishment is
that the retardation represents finite-energy transfer,
while the Hartree term corresponds to zero-energy transfer.
\par
\begin{figure}[t]
  \centering
  \includegraphics[width=1.0\linewidth]{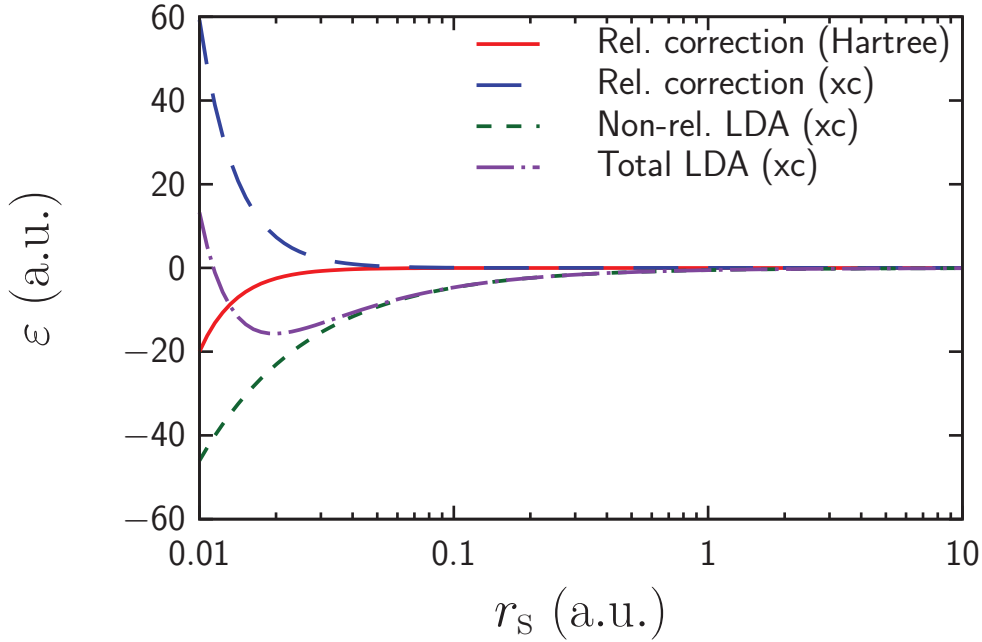}
  \caption{
    Relativistic corrections of the Hartree and the exchange--correlation (xc) energy densities
    shown with solid and long-dashed lines, respectively,
    as functions of the Wigner-Seitz radius $ r_{\urm{s}} $.
    For comparison, the Non-rel LDA exchange--correlation energy density in the PZ81 functional
    and the total LDA exchange--correlation energy density 
    are also shown with dashed and dash-dotted lines, respectively.}
  \label{fig:rel_func}
\end{figure}

The relativistic corrections of the Hartree and the exchange--correlation energy densities
are shown respectively with solid and long-dashed lines, 
as functions of the Wigner-Seitz radius $ r_{\urm{s}} $ in figure \ref{fig:rel_func}.
For comparison, the Non-rel LDA exchange--correlation energy density in the PZ81 functional
and the total LDA exchange--correlation energy density 
are also shown with dashed and dash-dotted lines, respectively.
The relativistic corrections are larger in the higher-density region,
i.e.,~the smaller $ r_{\urm{s}} $ region,
since the mean distance between two electrons $ r_{jk} $ in equation \eqref{eq:int} is smaller.
As $ r_{\urm{s}} $ increases, the relativistic corrections decrease
and reaches to zero in $ r_{\urm{s}} \to \infty $ limit.
Comparing the relativistic correction of the Hartree term with that of the exchange--correlation term,
the absolute value of the latter is around three times larger than that of the former.
\par
For calculation of isolated atoms, the spherical symmetry is assumed to the effective Kohn--Sham potential $ V_{\urm{KS}} $,
since the $ V_{\urm{ext}} $ has the spherical symmetry and is much stronger than the $ V_{\urm{int}} $.
In the one-body relativistic correction $ V'_1 $, the delta function is included,
and this term often causes numerical instability.
In order to avoid this problem,
the scalar-relativistic approximation
\cite{
  Koelling1977J.Phys.C10_3107}
\begin{equation}
  \label{eq:scalar}
  \hat{h}_{\urm{KS}}
  =
  - \frac{\hbar^2}{2M}
  \left[
    \laplace_r
    +
    \frac{l \left( l + 1 \right)}{r^2}
  \right]
  +
  V_{\urm{KS}} \left( r \right)
  -
  \frac{1}{4M^2 c^2}
  \frac{d V_{\urm{KS}}}{dr}
  \frac{d}{dr}
\end{equation}
is applied
to the single-particle Schr\"{o}dinger kinetic operator and the external potential,
$ \hat{t}^{\urm{S}} + V_{\urm{ext}} + V'_1 $,
where $ M $ is the energy-dependent effective mass
\begin{equation}
  \label{eq:effective_mass}
  M
  =
  m_e
  +
  \frac{\epsilon_j - V_{\urm{KS}}}{2c^2}, 
\end{equation}
$ \laplace_r $ is the radial component of the Laplacian
\begin{equation}
  \label{eq:laplace}
  \laplace_r
  =
  \frac{d^2}{dr^2}
  +
  \frac{2}{r}
  \frac{d}{dr},
\end{equation}
and $ l $ is the azimuthal quantum number.
With this approximation, the one-body relativistic effects $ V'_1 $ are included accurately.
Hence, the Kohn--Sham effective potential does not include $ V'_1 $ explicitly as 
\begin{equation}
  V_{\urm{KS}} \left( r \right)
  =
  V_{\urm{ext}} \left( r \right)
  + 
  \int
  \frac{\rho \left( r' \right)}{\left| \ve{r} - \ve{r}' \right|}
  \, d \ve{r}'
  + 
  V_{\urm{Hrel}} \left( r \right)
  + 
  V_{\urm{xc}} \left( r \right),
\end{equation}
where $ V_{\urm{xc}} $ is the exchange--correlation potential.
\par
It should be noted that if the spin-orbit interaction is added to the scalar-relativistic Hamiltonian, 
the eigenvalues and eigenfunctions of the Hamiltonian are exactly identical to those of the original Hamiltonian,
whereas the Foldy-Wouthuysen transformed Hamiltonian not
\cite{
  Liang2015Phys.Rep.570_1}.
\subsection{Spin-orbit Interaction}
\label{subsec:SO}
\par
The spin-orbit interaction, which is ignored with the scalar-relativistic approximation,
is treated as follows.
It is known that the spin-orbit interaction in electron systems is weak enough
\cite{
  Saue2011ChemPhysChem12_3077},
and hence, it is treated in the first-order perturbation theory.
The spin-orbit interaction in this scheme is
\begin{equation}
  \label{eq:SO}
  V_{\urm{SO}} \left( r \right)
  =
  - \frac{1}{4M^2 c^2}
  \frac{\kappa + 1}{r}
  \frac{d V_{\urm{KS}} \left( r \right)}{dr},
\end{equation}
where $ j = l \pm 1/2 $ is the total angular momentum.
Here, $ \kappa = + \left( j + 1/2 \right) $ for $ j = l + 1/2 $ and $ \kappa = - \left( j + 1/2 \right) $ for $ j = l - 1/2 $.
Correction of the single-particle energy due to the first-order perturbation theory for the spin-orbit interaction is
\begin{align}
  E_{n, l, j}^{\urm{SO}}
  & =
    \brakket{\psi_{n,l}}{V_{\urm{SO}}}{\psi_{n,l}} \notag \\
  & =
    - \frac{\kappa+1}{4}
    \int
    \frac{1}{\left[c + \left( \epsilon_{n,l} - V_{\urm{KS}} \right) / 2c \right]^2}
    \frac{d V_{\urm{KS}} \left( r \right)}{dr}
    \left|
    R_{n,l} \left( r \right)
    \right|^2
    r \, dr,
\end{align}
where $ \psi_{n,l} $ and $ \epsilon_{n,l} $ are the Kohn--Sham single-particle orbitals and energies, respectively, and $ R_{n,l} $ is the radial part of $ \psi_{n,l} $.
%
%
\section{Calculations and Discussion}
\label{sec:calc}
\par
In this section, calculation results given by the present scheme are shown.
Electronic properties of atomic systems are calculated as a benchmark.
The electron density is approximated to be the spherical symmetric.
\par
The scheme developed in this paper is called as `SRel-CB',
which is an abbreviation of the scalar-relativistic scheme with the Coulomb-Breit interaction.
The results are compared with those with the Non-rel and scalar-relativistic schemes without the Breit correction.
Here, the Non-rel scheme means the original Schr\"{o}dinger formalism.
The PZ81
\cite{
  Perdew1981Phys.Rev.B23_5048}
functional is used
as the exchange--correlation functional for the Non-rel and Scalar-rel schemes.
The ADPACK code \cite{
  ADPACK}
with implementing the SRel-CB scheme is used for the calculation.
\subsection{Radium}
\label{subsec:calc_1}
\par
The ground-state energy of a radium ($ Z = 88 $) atom is calculated.
The electronic configuration of the atom is $ \left[ \mathrm{Rn} \right] \, 7s^2 $.
Therefore, the spin-orbit interaction in the first-order perturbation theory does not affect the total ground-state energy and density.
\par
The electronic single-particle energy of the radium atom 
calculated in the SRel-CB scheme with and without the spin-orbit interaction (SO)
are shown in table \ref{tab:Ra_sp}.
For comparison, those calculated in the Non-rel and scalar-relativistic (Scalar-rel) schemes are also shown.
\par
We show in table \ref{tab:Ra_Etot} that
sum of the single-particle energies $ \sum_j \epsilon_j $,
the kinetic energy $ T_0 $,
the Hartree energy $ E_{\urm{H}} $,
the exchange--correlation energy $ E_{\urm{xc}} $,
the external potential energy $ E_{\urm{ext}} $, 
and the total energy $ E_{\urm{tot}} $ 
calculated in the SRel-CB, Non-rel, and Scalar-rel schemes.
For comparison, the ratio of each energy to that of the Non-rel scheme is also shown.
\par
\begin{figure}[t]
  \centering
  \includegraphics[width=1.0\linewidth]{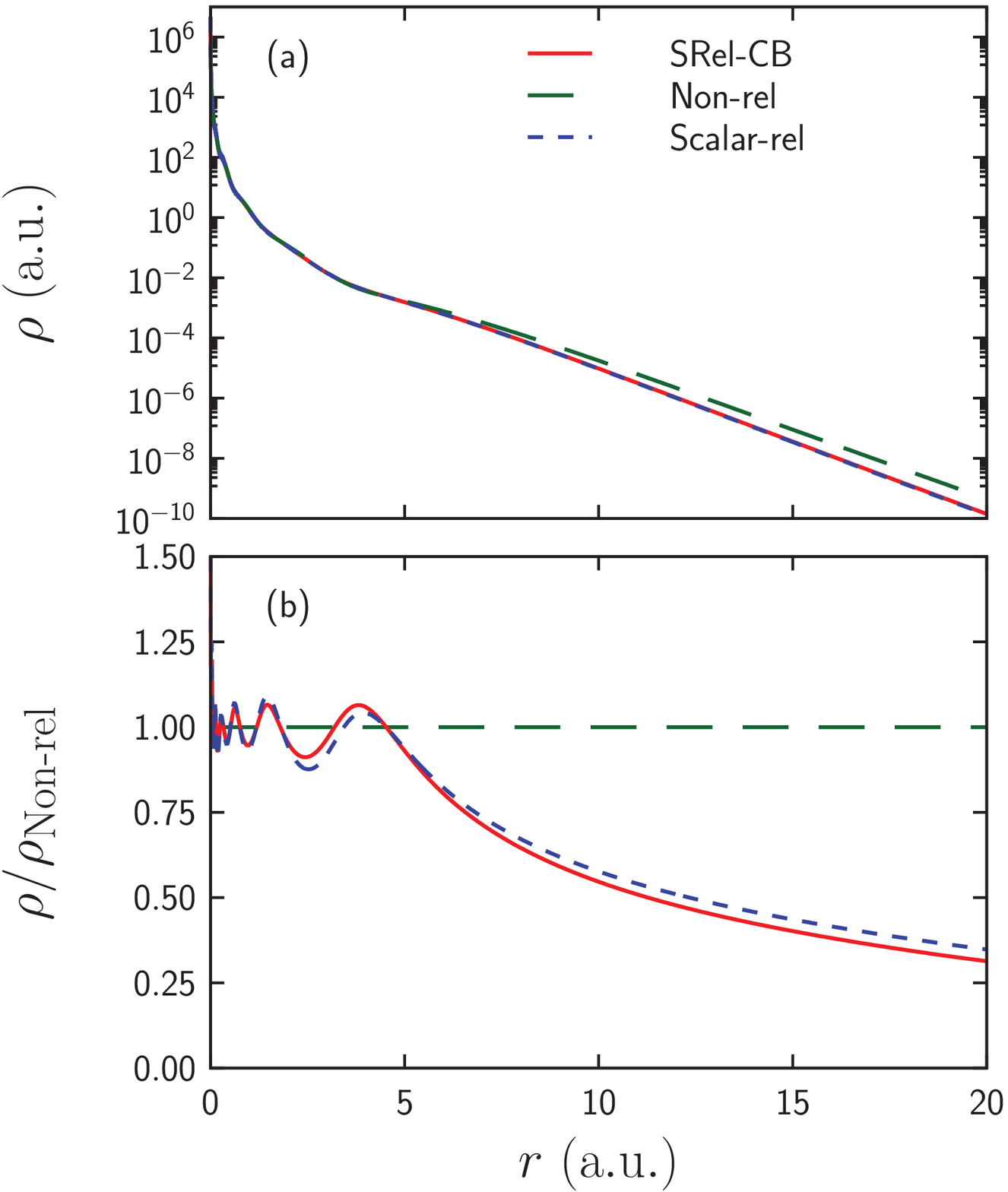}
  \caption{
    (a) Density distribution $ \rho \left( r \right) $ of radium atom calculated in the scalar-relativistic scheme with the Coulomb-Breit interaction (SRel-CB) shown with a solid line.
    (b) Ratio of density distribution in SRel-CB to that in the Non-rel scheme shown with a solid line.
    For comparison, those calculated in the Non-rel and scalar-relativistic (Scalar-rel) schemes
    (without the Breit correction)
    are also shown with long-dashed and dashed lines, respectively.}
  \label{fig:Ra_density}
\end{figure}

The density distribution $ \rho \left( r \right) $ calculated in the SRel-CB scheme is shown in figure \ref{fig:Ra_density}(a) as a solid line.
For comparison, those calculated in the Non-rel and Scalar-rel schemes are shown as long-dashed, dashed lines, respectively.
The ratio of density distribution to that in the Non-rel is shown in figure \ref{fig:Ra_density}(b).
\par
First, let us see the effect of the one-body correction $ V'_1 $
with the comparison between `Non-rel' and `Scalar-rel'
in tables \ref{tab:Ra_sp} and \ref{tab:Ra_Etot} and figure \ref{fig:Ra_density}.
Because $ V'_1 $ is the attractive,
it localizes the density more than that in the Non-rel scheme as shown in figure \ref{fig:Ra_density}.
Due to this localization, the external potential energy $ E_{\urm{ext}} $ and the kinetic energy $ T_0 $ are changed significantly as shown in table \ref{tab:Ra_Etot}.
The single-particle energies summarized in table \ref{tab:Ra_sp} indicate that
the $ s $ and $ p $ orbitals are bound more deeply due to the mass-velocity effect, 
while $ d $ and $ f $ orbitals are bound more shallowly in order to be orthonormal to $ s $ and $ p $ orbitals as known, e.g.,~in Refs.~\cite{
  Rose1978J.Phys.B11_1171,
  Pyykkoe1988Chem.Rev.88_563,
  Bartlett1998GoldBull.31_22}.
\par
Next, let us see the effect of the two-body correction $ V'_2 $
with the comparison between `Scalar-rel' and `SRel-CB'
in tables \ref{tab:Ra_sp} and \ref{tab:Ra_Etot} and figure \ref{fig:Ra_density}.
The effects of $ V'_2 $ is opposite of $ V'_1 $.
We can find in figure \ref{fig:Ra_density} that the density in the SRel-CB scheme is more delocalized than that in the Scalar-rel scheme.
The delocalization is understood as a consequence of the cancellation of non-relativistic and relativistic contributions to the exchange--correlation energy density $ \epsilon_{\urm{xc}} $.
Without $ V'_2 $, $ \epsilon_{\urm{xc}} $ has monotonically negative values near the ionic core as shown in figure \ref{fig:rel_func}.
The relativistic correction originating from $ V'_2 $ partially cancels this negative value,
which results in weakening of the electron attractive potential in the core region.
Note that this charge delocalization has little effect on the energies
$ \sum \epsilon_j $, $ T_0 $, $ E_{\urm{H}} $, and $ E_{\urm{ext}} $.
The change in $ E_{\urm{xc}} $ is found to be mainly due to the change of the functional form.
The single-particle energies summarized in table \ref{tab:Ra_sp} indicate that
the $ s $ and $ p $ orbitals are bound more shallowly.
Although the effect on $ d $ and $ f $ orbitals is non-monotonic,
we find a tendency that those with large principal quantum numbers bound more deeply.
Since the exchange--correlation energy contributes to the total energy less than $ 10 \, \% $,
even though that in the SRel-CB scheme is changed drastically from those in the other relativistic calculations, 
the density distribution in this work is almost the same.
More detailed discussion will be given in the next subsection.
\par
The spin-orbit interaction affects all the orbitals except the $ s $ orbitals.
Since its strength is often wholly comparable to $ V'_1 $,
once we incorporate this, the view on the $ V'_1 $ effects on the single-particle energies does not straightforwardly apply to the individual spin-orbit partners.
Note that the above discussion on $ V'_2 $ is still valid as seen from columns
`$ \text{Scalar-rel} + \text{SO} $' and `$ \text{SRel-CB} + \text{SO} $'.
\par
In short,
the $ s $ and $ p $ orbitals are bound more deeply due to $ V'_1 $,
and more shallowly due to $ V'_2 $.
In total, they are bound more deeply since the effect of $ V'_1 $ is larger than that of $ V'_2 $.
The $ d $ and $ f $ orbitals tend to show opposite behavior from the $ s $ and $ p $ orbitals.
\begin{table}[t]
  \centering
  \caption{
    Single-particle energies of radium calculated in the scalar-relativistic scheme with the Coulomb-Breit interaction without and with the spin-orbit interaction as `SRel-CB' and `SRel-CB + SO'.
    For comparison, those calculated in Non-rel and Scalar-rel schemes with and without the spin-orbit interaction are also shown.
    The spin-orbit interaction is considered as the first-order perturbation theory discussed in section \ref{subsec:SO}.
    In the fourth and sixth columns, 
    energies of the spin-orbit partners with $ j = l - 1/2 $ and $ j = l + 1/2 $
    are shown on the upper and lower rows of each $ n $ and $ l $, respectively.
    All units are in the Hartree atomic unit.}
  \label{tab:Ra_sp}
  \begin{indented}
  \item[]
    \begin{tabular}{@{}lD{.}{.}{5}D{.}{.}{5}D{.}{.}{5}D{.}{.}{5}D{.}{.}{5}D{.}{.}{5}D{.}{.}{5}}
      \br
      Orbitals & \multicolumn{1}{c}{Non-rel} & \multicolumn{1}{c}{Scalar-rel} & \multicolumn{1}{c}{Scalar-rel+SO} & \multicolumn{1}{c}{SRel-CB} & \multicolumn{1}{c}{SRel-CB + SO} \\ 
      \mr
      $ 1s $ & -3362.71476  & -3821.91003  & -3821.91003 & -3778.36010 & -3778.36010 \\
      $ 2s $ &  -577.09970  &  -702.12514  &  -702.12514 &  -695.88354 &  -695.88354 \\
      $ 2p $ &  -557.51465  &  -591.77045  &  -654.11813 &  -589.63834 &  -651.33399 \\
               &            &              &  -560.59661 &             &  -558.79051 \\
      $ 3s $ &  -142.63234  &  -174.15965  &  -174.15965 &  -172.79380 &  -172.79380 \\
      $ 3p $ &  -133.12385  &  -143.17134  &  -157.39457 &  -142.75177 &  -156.83880 \\
               &            &              &  -136.05972 &             &  -135.70826 \\
      $ 3d $ &  -115.30711  &  -114.39221  &  -117.55364 &  -114.28743 &  -117.43874 \\
               &            &              &  -112.28460 &             &  -112.18655 \\
      $ 4s $ &   -34.52561  &   -42.70358  &   -42.70358 &   -42.36599 &   -42.36599 \\
      $ 4p $ &   -30.22136  &   -32.56286  &   -36.17526 &   -32.47980 &   -36.05903 \\
               &            &              &   -30.75667 &             &   -30.69018 \\
      $ 4d $ &   -22.20826  &   -21.73159  &   -22.45257 &   -21.73097 &   -22.45008 \\
               &            &              &   -21.25904 &             &   -21.25157 \\
      $ 4f $ &   -11.18118  &   -10.02204  &   -10.19512 &   -10.05025 &   -10.22341 \\
               &            &              &    -9.89223 &             &    -9.92038 \\
      $ 5s $ &    -7.13875  &    -8.90000  &    -8.90000 &    -8.82881 &    -8.82881 \\
      $ 5p $ &    -5.54683  &    -5.86012  &    -6.65487 &    -5.85085 &    -6.63892 \\
               &            &              &    -5.46275 &             &    -5.45682 \\
      $ 5d $ &    -2.81942  &    -2.54383  &    -2.66536 &    -2.55132 &    -2.67273 \\
               &            &              &    -2.46282 &             &    -2.47037 \\
      $ 6s $ &    -1.05108  &    -1.29137  &    -1.29137 &    -1.28132 &    -1.28132 \\
      $ 6p $ &    -0.634553 &    -0.613685 &   -0.726722 &   -0.614691 &   -0.726969 \\
               &            &              &   -0.557166 &             &   -0.558552 \\
      $ 7s $ &    -0.113918 &    -0.125796 &   -0.125796 &   -0.125299 &   -0.125299 \\
      \br
    \end{tabular}
  \end{indented}
\end{table}
\begin{table}[t]
  \centering
  \caption{
    Sum of the single-particle energy $ \sum_j \epsilon_j $,
    the kinetic energy $ T_0 $,
    the Hartree energy $ E_{\urm{H}} $,
    the exchange--correlation energy $ E_{\urm{xc}} $,
    the external potential energy $ E_{\urm{ext}} $, 
    and the total energy $ E_{\urm{tot}} $ of radium atom
    calculated in the Non-rel, Scalar-rel, and SRel-CB schemes.
    In order to compare, the normalized energies where those in the Non-rel scheme are normalized in $ 100.0 \, \% $ are also shown.
    All units for the energies are in the Hartree atomic unit.}
  \label{tab:Ra_Etot}
  \begin{indented}
  \item[]
    \begin{tabular}{@{}lD{.}{.}{5}D{.}{.}{5}D{.}{.}{5}}
      \br
      & \multicolumn{1}{c}{Non-rel} & \multicolumn{1}{c}{Scalar-rel} & \multicolumn{1}{c}{SRel-CB} \\
      \mr
      \multicolumn{1}{c}{$ \sum_j \epsilon_j $}      & -14172.68427 & -15673.28668 & -15553.69107 \\
      \multicolumn{1}{c}{$ T_0 $}                    &  23081.25534 &  29327.98912 &  29009.39740 \\
      \multicolumn{1}{c}{$ E_{\urm{H}} $}            &   9045.58330 &   9420.50692 &   9372.61666 \\
      \multicolumn{1}{c}{$ E_{\urm{xc}} $}           &   -395.67257 &   -425.64473 &   -325.19714 \\
      \multicolumn{1}{c}{$ E_{\urm{ext}} $}          & -54819.79719 & -63277.04574 & -62943.31742 \\
      \multicolumn{1}{c}{$ E_{\urm{tot}} $}          & -23088.63112 & -24954.19443 & -24886.50050 \\
      \mr
      \multicolumn{1}{c}{$ \sum_j \epsilon_j $}      & 100.0 & 110.58799 & 109.74414 \\
      \multicolumn{1}{c}{$ T_0 $}                    & 100.0 & 127.06410 & 125.68379 \\
      \multicolumn{1}{c}{$ E_{\urm{H}} $}            & 100.0 & 104.14483 & 103.61539 \\
      \multicolumn{1}{c}{$ E_{\urm{xc}} $}           & 100.0 & 107.57499 & 82.18847 \\
      \multicolumn{1}{c}{$ E_{\urm{ext}} $}          & 100.0 & 115.42736 & 114.81859 \\
      \multicolumn{1}{c}{$ E_{\urm{tot}} $}          & 100.0 & 108.08001 & 107.78684 \\
      \br
    \end{tabular}
  \end{indented}  
\end{table}
\subsection{Groups 1, 2, and 18 Elements}
\label{subsec:calc_2}
\par
In order to discuss the systematic behavior of relativistic effects in this scheme,
properties of all the groups 1, 2, and 18 atoms are calculated.
We do not address the hydrogen atom
since it has only one electron, and therefore, the exchange--correlation term is zero.
All the group 18 atoms are closed shell and all the group 1 and 2 atoms are closed-shell plus $ s $ electrons.
Therefore, the spin-orbit interaction does not affect the total energies in these atoms in the first-order perturbation theory.
\par
The total energies calculated in the Non-rel, Scalar-rel, and SRel-CB schemes
are shown in table \ref{tab:systematic_total}.
It is seen that the relativistic effects of $ V'_2 $ are 
non-negligible in heavier atoms as well as those of $ V'_1 $,
and the former effects for the total energy are opposite to the latter.
\par
\begin{figure}[t]
  \centering
  \includegraphics[width=1.0\linewidth]{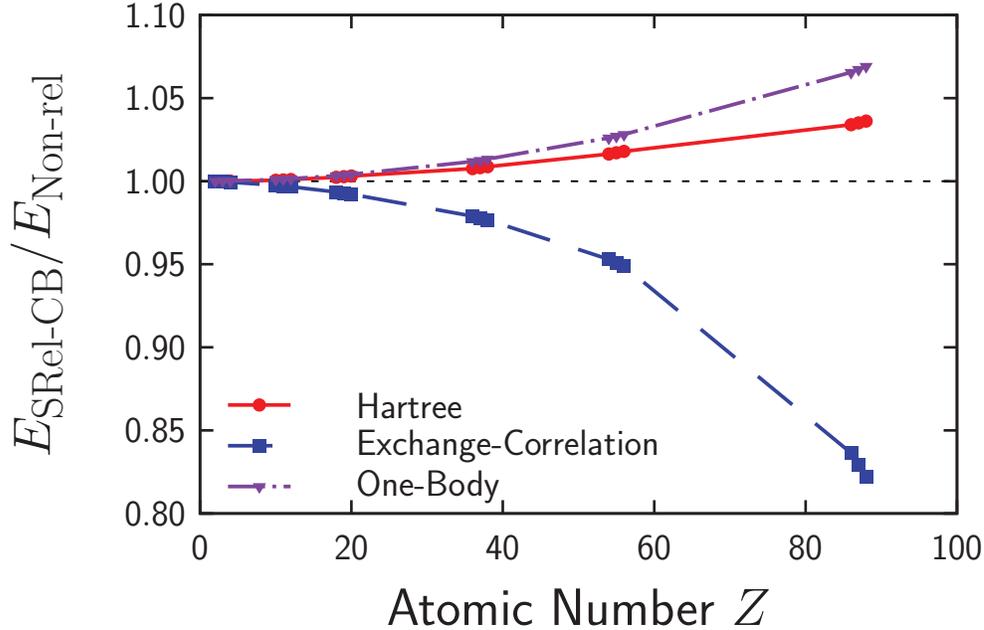}
  \caption{
    Ratio of the energies calculated in the SRel-CB scheme to those in the Non-rel one.
    The Hartree energy $ E_{\urm{H}} $, exchange--correlation energy $ E_{\urm{xc}} $,
    and energy from the one-body term $ T_0 + E_{\urm{ext}} $ are shown in solid, long-dashed, and ash-dotted lines, respectively.}
  \label{fig:ratio_schr}
\end{figure}
\begin{figure}[t]
  \centering
  \includegraphics[width=1.0\linewidth]{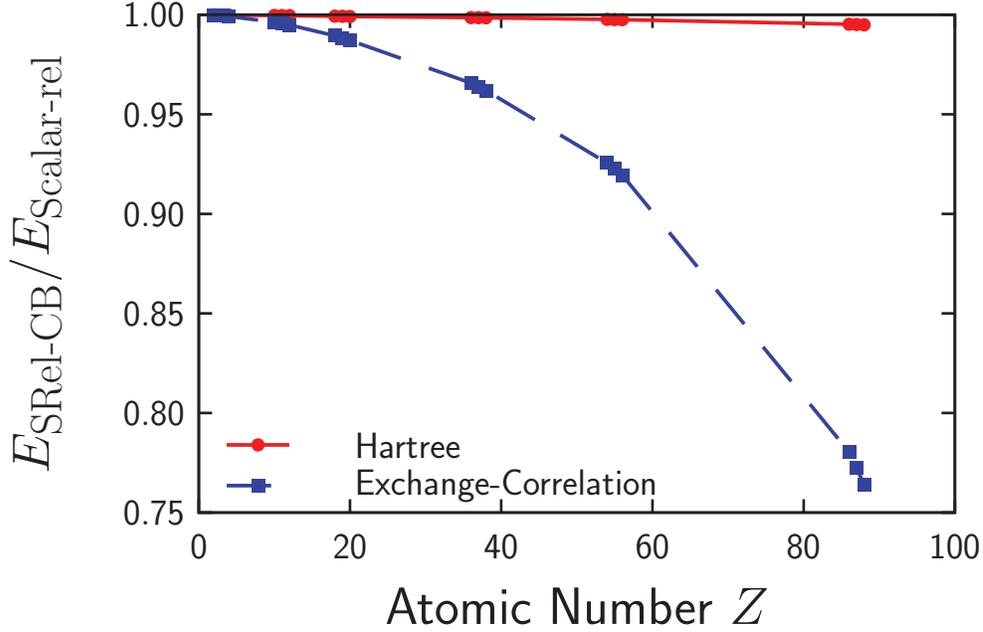}
  \caption{
    Same as figure \ref{fig:ratio_schr} but compared with the Scalar-rel scheme.
    Since the one-body operator in this work is the same as that in the scalar-relativistic calculation,
    energy from the one-body term $ T_0 + E_{\urm{ext}} $ is not shown.}
  \label{fig:ratio_scalar}
\end{figure}

We here analyze the contributions of the Hartree $ E_{\urm{H}} $ and exchange--correlation $ E_{\urm{xc}} $ energies and energy from the one-body term $ T_0 + E_{\urm{ext}} $ separately.
The ratios of these values in the SRel-CB scheme, $ E_{\urm{SRel-CB}} $, to those in the Non-rel scheme, $ E_{\urm{non-rel}} $, are shown in figure \ref{fig:ratio_schr},
and ratios to those in the Scalar-rel, $ E_{\urm{Scalar-rel}} $, are shown in figure \ref{fig:ratio_scalar}.
Those for the Hartree energy $ E_{\urm{H}} $, exchange--correlation energy $ E_{\urm{xc}} $,
and energy from the one-body term $ T_0 + E_{\urm{ext}} $ are shown in solid, long-dashed, and dash-dotted lines, respectively.
Since the one-body operator in the SRel-CB scheme is the same as that in the Scalar-rel scheme,
the ratio of the one-body term $ T_0 + E_{\urm{ext}} $ is not shown.
\par
The relativistic correction to the external potential $ V'_1 $,
which is the attractive force,
makes the external attractive potential stronger,
whereas that to the interaction $ V'_2 $,
which is also the attractive force,
makes the repulsive interaction smaller.
Thus, the relativistic effects make the energy due to the potential, $ E_{\urm{ext}} $, larger,
while that makes the energies due to the interaction, $ E_{\urm{H}} $ and $ E_{\urm{xc}} $, weaker,
as shown in figure \ref{fig:ratio_schr}.
However, a lot of effects,
such as the change of the Kohn--Sham orbitals,
are entangled to each other in the self-consistent step
and eventually $ E_{\urm{H}} $ in the SRel-CB scheme is larger than that in the Non-rel scheme.
\par
Between $ E_{\urm{H}} $ and $ E_{\urm{xc}} $, $ E_{\urm{SRel-CB}} / E_{\urm{Scalar-rel}} $ for $ E_{\urm{H}} $ is smaller than $ E_{\urm{xc}} $.
The interaction includes the finite-light-speed effect as well as effects coming from the Dirac equation,
whereas the finite-light-speed effect of the Hartree term vanishes.
As a result, the relativistic correction of the Hartree energy is smaller than that of the exchange--correlation energy, 
as also discussed in section \ref{subsec:finite}.
Since the absolute value of the Non-rel Hartree term is larger than that of the Non-rel exchange--correlation term,
finally, the ratios $ E_{\urm{Scalar-rel}} / E_{\urm{Non-rel}} $ and $ E_{\urm{SRel-CB}} / E_{\urm{Non-rel}} $ of the Hartree energy are further smaller than those of the exchange--correlation energy
as shown in figure \ref{fig:ratio_scalar}.
Therefore, it can be concluded that 
the finite-light-speed correction is less significant than effects coming from the Dirac equation
for the electron-electron interaction.
\par
One can also find that the relativistic corrections are more significant in larger $ Z $.
According to figure \ref{fig:rel_func}, the relativistic corrections are significant in $ r_{\urm{s}} \lesssim 0.1 \, \mathrm{a.u.} $ region.
As $ Z $ increases, the density also increases and notably, 
the region with $ r_{\urm{s}} \lesssim 0.1 \, \mathrm{a.u.} $ extends.
In consequence, the relativistic effects in larger $ Z $ are more significant,
as shown in figures \ref{fig:ratio_schr} and \ref{fig:ratio_scalar}.
\begin{table*}[t]
  \centering
  \caption{
    Total energies for selected atoms in the SRel-CB scheme.
    For comparison, total energies calculated in the Non-rel and Scalar-rel schemes are also shown.
    All units are in the Hartree atomic unit.}
  \label{tab:systematic_total}
  \begin{indented}
  \item[]
    \begin{tabular}{@{}lrD{.}{.}{5}D{.}{.}{5}D{.}{.}{5}}
      \br
      Atoms & $ Z $ & \multicolumn{1}{c}{Non-rel} & \multicolumn{1}{c}{Scalar-rel} & \multicolumn{1}{c}{SRel-CB} \\
      \mr
      Helium  &    2 &     -2.83435 &     -2.83448 &     -2.83439 \\
      Lithium &    3 &     -7.33420 &     -7.33499 &     -7.33457 \\
      Beryllium &  4 &    -14.44637 &    -14.44920 &    -14.44800 \\
      Neon &      10 &   -128.22811 &   -128.37290 &   -128.34637 \\
      Sodium &    11 &   -161.43435 &   -161.65367 &   -161.61680 \\
      Magnesium & 12 &   -199.13369 &   -199.45449 &   -199.40493 \\
      Argon &     18 &   -525.93971 &   -527.80802 &   -527.61130 \\
      Potassium & 19 &   -598.19357 &   -600.55712 &   -600.32052 \\
      Calcium &   20 &   -675.73508 &   -678.68967 &   -678.40788 \\
      Krypton &   36 &  -2750.13629 &  -2786.82081 &  -2784.74430 \\
      Rubidium &  37 &  -2936.32553 &  -2977.62068 &  -2975.33556 \\
      Strontium & 38 &  -3129.44131 &  -3175.78666 &  -3173.27751 \\
      Xenon &     54 &  -7228.83884 &  -7441.14722 &  -7432.27164 \\
      Caesium &   55 &  -7550.54003 &  -7780.59333 &  -7771.08692 \\
      Barium &    56 &  -7880.09328 &  -8129.04311 &  -8118.86879 \\
      Radon &     86 & -21861.29405 & -23538.40147 & -23477.93363 \\
      Francium &  87 & -22470.26526 & -24239.48311 & -24175.50669 \\
      Radium &    88 & -23088.63112 & -24954.19443 & -24886.50050 \\
      \br
    \end{tabular}
  \end{indented}
\end{table*}
\subsection{Lawrencium and Lutetium}
\label{subsec:calc_Lr}
\par
We compare the energies of two cases of the electronic configuration of lawrencium atoms,
where one valence electron occupies
the $ 6d $ orbital
($ \left[ \mathrm{Rn} \right] \, 5f^{14} \, 6d^1 \, 7s^2 $) 
or
occupies the $ 7p $ orbital
($ \left[ \mathrm{Rn} \right] \, 5f^{14} \, 7s^2 \, 7p^1 $).
For comparison, those of lutetium atoms are also calculated,
where one valence electron occupies
the $ 5d $ orbital
($ \left[ \mathrm{Xe} \right] \, 4f^{14} \, 5d^1 \, 6s^2 $) 
or
occupies the $ 6p $ orbital
($ \left[ \mathrm{Xe} \right] \, 4f^{14} \, 6s^2 \, 6p^1 $).
The non-spherical modification in $ V_{\urm{KS}} $
is ignored for simplicity.
In principle, both the lutetium and lawrencium are open-shell atoms,
and thus the effective potential $ V_{\urm{KS}} $ may be non-spherical.
Since there is strong spherical (central) external potential caused by the atomic nucleus,
it is assumed that the single-particle energies and total energies are scarcely affected by the non-sphericity of $ V_{\urm{KS}} $.
\par
The energies calculated with the above-mentioned configuration are shown in table \ref{tab:Lr_Etot}.
All energies are calculated in the Non-rel, Scalar-rel, and SRel-CB schemes.
The smaller values for the respective approximations in each atom are shown with the bold font.
\par
On the one hand,
in lutetium atoms,
the electron prefers to occupy the $ d $ orbital rather than the $ p $ orbital
in all schemes.
On the other hand,
in lawrencium atoms,
the electron prefers to occupy the $ p $ orbital rather than the $ d $ orbital
in the SRel-CB scheme,
whereas it still prefers the $ d $ orbital in the Non-rel scheme
like the lutetium case.
In addition, the electron is unbound for lawrencium atoms
in the Scalar-rel scheme.
We can understand the dependence of the electronic configuration in lawrencium atoms on the level of the approximation with  the single-particle orbitals discussed in section \ref{subsec:calc_1}.
Due to the relativistic corrections $ V'_1 $ and $ V'_2 $,
the $ 6d $ orbital may be bound shallower,
while the $ 7p $ orbital may be bound deeper,
compared with the results in the Non-rel scheme.
As a result, the $ 6d $ orbital may becomes unbound, while the $ 7p $ orbital is still kept bound.
\par
The present results suggest that a valence electron occupies $ p $ orbitals in lawrencium atoms,
where the finite-light-speed correction to the Coulomb interaction has a crucial role.
The occupation of $ p $ orbital could be the origin of the anomalous behavior of the lawrencium \cite{Sato2015Nature520_209,IUPAC}.
\begin{table}[t]
  \centering
  \caption{
    Assumed electronic configurations and total energies for lutetium and lawrencium atoms in the Non-rel, Scalar-rel, and SRel-CB schemes.
    Two types of electronic configurations are considered.
    The lower energies in each calculation are shown with the bold font.
    All units are in the Hartree atomic unit.}
  \label{tab:Lr_Etot}
  \begin{indented}
  \item[]
    \begin{tabular}{@{}llccc}
      \br
      Atoms & Configuration & \multicolumn{1}{c}{Non-rel} & \multicolumn{1}{c}{Scalar-rel} & \multicolumn{1}{c}{SRel-CB} \\
      \mr
      $ \mathrm{Lu} $ & $ \left[ \mathrm{Xe} \right] \, 4f^{14} \, 5d^1 \, 6s^2 $ & $ \mathbf{-13848.19912} $ & $ \mathbf{-14533.23001} $ & $ \mathbf{-14527.46807} $ \\
            & $ \left[ \mathrm{Xe} \right] \, 4f^{14} \, 6s^2 \, 6p^1 $ & $ -13848.12376 $ & $ -14533.19019 $ & $ -14527.42700 $ \\ \hline
      $ \mathrm{Lr} $ & $ \left[ \mathrm{Rn} \right] \, 5f^{14} \, 6d^1 \, 7s^2 $ & $ \mathbf{-33551.48205} $ & \multicolumn{1}{c}{Unbound} & \multicolumn{1}{c}{Unbound} \\
            & $ \left[ \mathrm{Rn} \right] \, 5f^{14} \, 7s^2 \, 7p^1 $ & $ -33551.38274 $ & \multicolumn{1}{c}{Unbound} & $ \mathbf{-37331.31054} $ \\
      \br
    \end{tabular}
  \end{indented}
\end{table}
%
%
\section{Conclusion}
\label{sec:conc}
\par
In this paper,
a non-relativistic-reduction-based approach to the local density approximation for density functional theory (DFT) with the relativistic correction up to
$ O \left( \left( Z \alpha \right)^2 \right) \sim O \left( 1/c^2 \right) $ has been formulated.
Since in this paper the Hartree term with the Breit correction is derived,
DFT becomes able to treat the finite-light-speed correction to the Coulomb interaction.
Note that the Hamiltonian used in this paper is the same as
the Hamiltonian previously used
in wave-function methods to discuss the electronic structure of super-heavy elements,
and the way to construct DFT scheme from the Hamiltonian is the same as the case of Non-rel LDA.
\par
There exist two relativistic corrections:
one is for one-body potential $ V'_1 $ and
the other is for two-body interaction $ V'_2 $.
The former is stronger than the latter
and gives opposite contribution than the latter for the Kohn--Sham potential.
In total, relativistic effects make
the $ s $ and $ p $ orbitals bound more deeply and $ d $ and $ f $ orbitals more shallowly.
Although $ V'_2 $ makes the $ s $ and $ p $ orbitals bound more shallowly and $ d $ and $ f $ orbitals bound more deeply,
the effect of $ V'_1 $ dominates.
\par
According to the calculation in this work, DFT with the finite-light-speed correction, the outer-most electron of lawrencium atoms occupies the $ p $ orbital whereas that of lutetium atoms occupies the $ d $ orbital.
This different electronic configuration may cause the anomaly of the ionization energy of lawrencium atoms.
This result is consistent with the previous works calculated by wave-function theory
\cite{
  Desclaux1980J.Phys.France41_943,
  Eliav1995Phys.Rev.A52_291,
  Cao2003J.Chem.Phys.118_487,
  Eliav2015Nucl.Phys.A944_518,
  Pershina2015Nucl.Phys.A944_578}, 
while the computational cost is lower than those of previous works.
It should be noted that the $ p $-block elements are defined as those whose outer-most electrons occupy the $ p $ orbitals \cite{Cotton1995_JohnWiley&Sons}. 
Lawrencium thus belongs to the $ p $ block, not $ d $ according to the definition. 
It seems, however, more appropriate to regard that the concept of `block' becomes ambiguous for the heavy elements, as the electronic configuration of lawrencium is almost the same as that of lutetium, apart from the only difference of the outer-most electron. 
Reconsideration of the classification appropriate for super-heavy elements would be mandatory.
\par
For the atomic systems, the accurate wave-function theory such as the CI and CC methods are feasible,
but they become impractical for molecular and solid systems compared with DFT.
Since the relativistic effects, as the same way as the wave-function methods,
are now implemented to DFT in this work and
it provides consistent results with those by the wave-function methods,
with which reliable calculation of properties of molecules and solids of super-heavy elements is expected to be feasible.
These complementary methods may help to understand and predict the atomic properties of the super- and hyper-heavy elements.
In the future, theoretical prediction of the periodic table of the elements may be attained with these complementary methods.
\par
In addition, precise calculation and measurement of the super- and hyper-heavy elements will help to test the QED \cite{Indelicato2007Eur.Phys.J.D45_155,Indelicato2013Phys.Rev.A87_022501} and the electric dipole moment of electrons and atomic nuclei, which is related to $ CP $ and $ T $ symmetries \cite{Pospelov2005Ann.Phys.318_119,Yamanaka2017Eur.Phys.J.A53_54},
as well as properties of the atoms itself.
\par
It is known that some properties of solids are better reproduced by GGA instead of LDA \cite{Asada1992Phys.Rev.B46_13599}.
So far only the relativistic version of the B88 exchange functional \cite{Engel1996Phys.Rev.A53_1367} has been known,
which includes some empirical parameters. 
Thus, non-empirical relativistic exchange--correlation functionals within GGA is interesting.
\par
Relativistic effects for the spin-polarized systems are also interesting.
In the two-body correction $ V'_2 $,
the spin-orbit and spin-spin interactions between two electrons exist,
while this term vanishes in the spin-unpolarized systems.
This effect has never been considered and this may give rise to non-trivial phenomena.
Also, the Hartree energy due to the retardation term $ V'_2 $, 
which is zero in the time-reversal symmetric case, can be nonzero.
In order to consider these effects in the calculation of solids,
construction of the pseudopotential is also required \cite{Martin2004_CambridgeUniversityPress}.
%
%
\ack
\par
T.N.~would like to thank 
Hiromitsu Haba,
Kouichi Hagino,
Naoyuki Itagaki, 
Yasuhiro Sakemi, 
Tetsuya K.~Sato,
and Kenichi Yoshida
for fruitful discussions and variable comments.
T.N.~and H.L.~would like to thank the RIKEN iTHEMS program,
the JSPS-NSFC Bilateral Program for Joint Research Project on Nuclear mass and life for unravelling mysteries of the $ r $-process,
and the RIKEN Pioneering Project: Evolution of Matter in the Universe.
T.N.~would also like to thank the visitor program of the Yukawa Institute for Theoretical Physics, Kyoto University
and the JSPS Grant-in-Aid for JSPS Fellows under Grant No.~19J20543.
H.L.~acknowledges the JSPS Grant-in-Aid for Early-Career Scientists under Grant No.~18K13549.
The numerical calculations have been performed on cluster computers at Department of Physics, The University of Tokyo and the RIKEN iTHEMS program.
%
\appendix
%
\section{Two Relativistic Correction for the Exchange--Correlation Functionals}
\label{sec:app_xc}
\par
The LDA form derived by Kenny \textit{et al\/.}~\cite{Kenny1996Phys.Rev.Lett.77_1099} is used in this paper
to consider the Breit correction for the exchange--correlation functional $ E_{\urm{xc}} $.
This form has been constructed in the same way as the Non-rel LDA exchange--correlation functional, PZ81 \cite{Perdew1981Phys.Rev.B23_5048},
while the Coulomb-Breit interaction
is used for the electron-electron interaction $ V_{\urm{int}} $ 
instead of the Coulomb interaction.
\par
There is another relativistic correction for the exchange functional in LDA derived by MacDonald and Vosko \cite{MacDonald1979J.Phys.C12_2977}.
This functional is constructed in the same way as the Non-rel LDA exchange functional, i.e.,~the Hartree--Fock--Slater approximation as
\begin{equation}
  \epsilon_{\urm{x}} \left( \rho \right)
  = 
  - \frac{3}{4}
  \left( \frac{3}{\pi} \right)^{1/3}
  \rho^{1/3} 
  \left[
    1 
    -
    \frac{2}{3}
    \frac{\left( 3 \pi^2 \rho \right)^{2/3}}{c^2}
  \right].
\end{equation}
The relativistic correction to the correlation part is not considered in this functional.
\par
In this appendix, the above-mentioned two relativistic corrections for $ E_{\urm{xc}} $ are discussed.
The relativistic corrections derived by Kenny \textit{et al\/.}~\cite{Kenny1996Phys.Rev.Lett.77_1099} 
and by MacDonald and Vosko \cite{MacDonald1979J.Phys.C12_2977}
are referred to as `LDA-RK' and `LDA-RMV', respectively.
\par
\begin{figure}[t]
  \centering
  \includegraphics[width=1.0\linewidth]{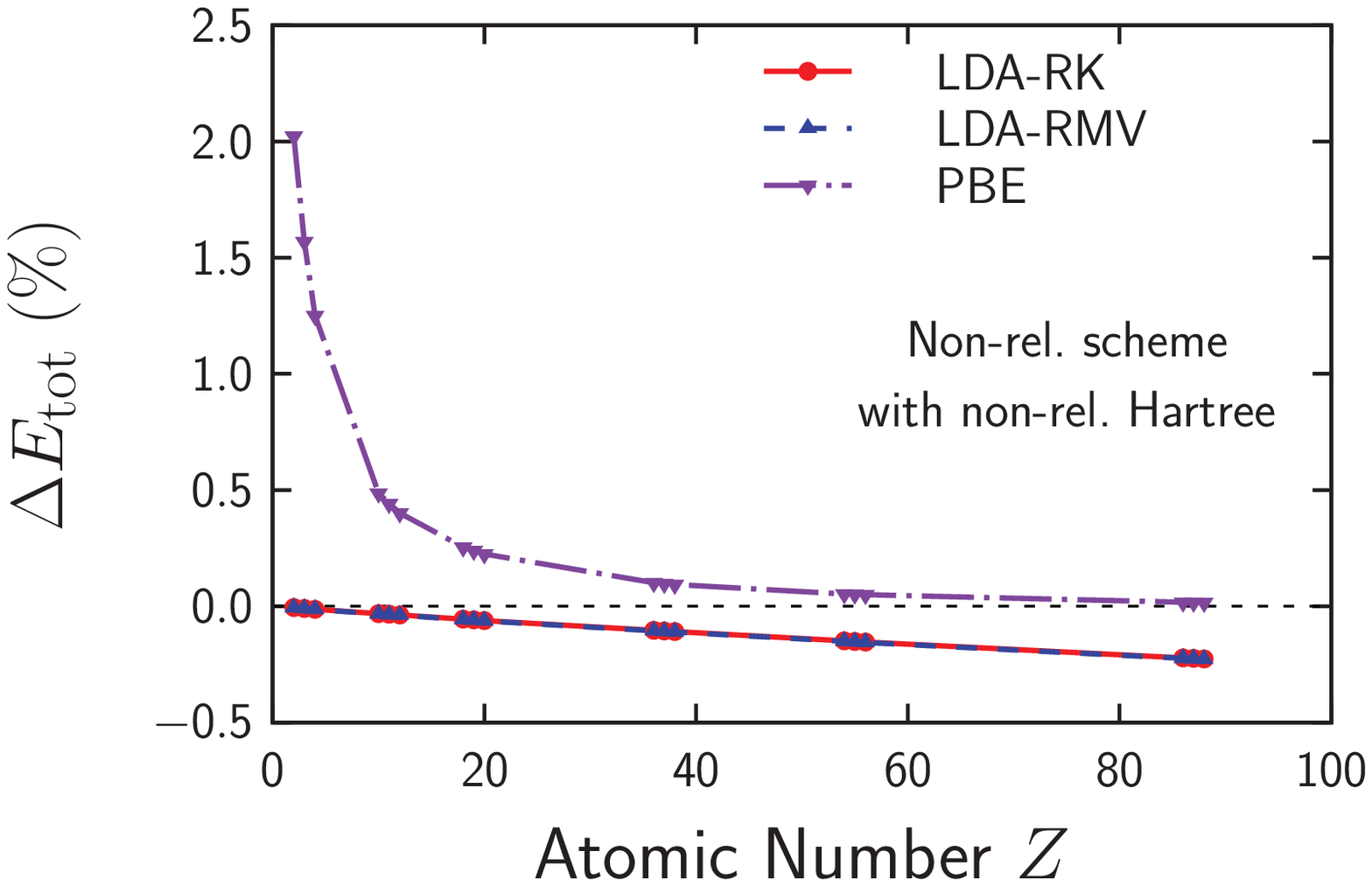}
  \caption{
    Ratio of the total energy calculated with the LDA-RK, LDA-RMV, and PBE functionals to that calculated with the PZ81, $ \Delta E_{\urm{tot}} $ 
    shown in solid, dashed, and dash-dotted lines, respectively.}
  \label{fig:ratio_app}
\end{figure}

In table \ref{tab:app_data}, the total energies calculated in the non-relativistic and relativistic LDA exchange--correlation functional are shown,
where the PZ81 functional is used for the non-relativistic functional, 
while the LDA-RK and LDA-RMV functionals are used for the relativistic functional.
For comparison, the result with the Non-rel GGA exchange--correlation functional is also shown,
where the PBE functional \cite{Perdew1996Phys.Rev.Lett.77_3865} is used.
For the kinetic term and the Hartree term, the Non-rel scheme is used.
In figure \ref{fig:ratio_app}, ratio of the total energy calculated with LDA-RK, LDA-RMV, and PBE to that calculated with PZ81
\begin{equation}
  \Delta E_{\urm{tot}} 
  =
  \frac{E_{\urm{tot}} - E_{\urm{tot}}^{\urm{PZ81}}}{E_{\urm{tot}}}
\end{equation}
are shown in solid, dashed, and dash-dotted lines, respectively.
\par
On the one hand, LDA-RK includes the relativistic correction for the correlation energy 
together with that for the relativistic exchange energy.
On the other hand, LDA-RMV includes only the correction for the exchange energy.
Two functionals give almost the same results.
Therefore, the relativistic correction for the correlation term is negligible.
\par
In addition, even in $ Z \simeq 40 $ region, the relativistic correction and the gradient correction for the total energy are comparable while the signs of $ \Delta E_{\urm{tot}} $ are opposite to each other.
In $ Z > 50 $ region, the relativistic correction for the total energy is larger than the gradient correction.
This example implies that the impact of the relativistic correction can be as significant as the gradient correction in a wide range of the systems.
\begin{table}[t]
  \centering
  \caption{
    Total energies calculated in the non-relativistic and relativistic LDA exchange--correlation functionals.
    For the Non-rel functional, the PZ81 functional is used,
    while for the relativistic functional, the LDA-RK and the LDA-RMV functionals are used.
    For comparison the Non-rel GGA exchange--correlation functional is also shown,
    where the PBE functional \cite{Perdew1996Phys.Rev.Lett.77_3865} is used.
    For the kinetic term and the Hartree term, the Non-rel scheme is used.}
  \label{tab:app_data}
  \begin{indented}
  \item[]
    \begin{tabular}{@{}lrD{.}{.}{5}D{.}{.}{5}D{.}{.}{5}D{.}{.}{5}}
      \br
      \multicolumn{1}{l}{Atoms} & \multicolumn{1}{c}{$ Z $} & \multicolumn{1}{c}{PZ81} & \multicolumn{1}{c}{PBE} & \multicolumn{1}{c}{LDA-RK} & \multicolumn{1}{c}{LDA-RMV} \\
      \mr
      Helium & 2     & -2.83435 & -2.89288 & -2.83419 & -2.83417 \\
      Lithium & 3    & -7.33420 & -7.45114 & -7.33353 & -7.33345 \\
      Beryllium & 4  & -14.44637 & -14.62934 & -14.44450 & -14.44433 \\
      Neon & 10      & -128.22811 & -128.85570 & -128.18769 & -128.18595 \\
      Sodium & 11    & -161.43435 & -162.15032 & -161.37850 & -161.37630 \\
      Magnesium & 12 & -199.13369 & -199.93645 & -199.05876 & -199.05601 \\
      Argon & 18     & -525.93971 & -527.28209 & -525.64756 & -525.63985 \\
      Potassium & 19 & -598.19357 & -599.62967 & -597.84352 & -597.83468 \\
      Calcium & 20   & -675.73508 & -677.26086 & -675.31933 & -675.30929 \\
      Krypton & 36   & -2750.13629 & -2752.92551 & -2747.28274 & -2747.23853 \\
      Rubidium & 37  & -2936.32553 & -2939.18962 & -2933.20263 & -2933.15523 \\
      Strontium & 38 & -3129.44131 & -3132.37553 & -3126.03205 & -3125.98133 \\
      Xenon & 54     & -7228.83884 & -7232.68662 & -7218.08285 & -7217.95898 \\
      Caesium & 55   & -7550.54003 & -7554.42904 & -7539.11954 & -7538.98974 \\
      Barium & 56    & -7880.09328 & -7884.01939 & -7867.98002 & -7867.84414 \\
      Radon & 86     & -21861.29405 & -21865.09427 & -21812.86582 & -21812.45816 \\
      Francium & 87  & -22470.26526 & -22474.02687 & -22419.98747 & -22419.56745 \\
      Radium & 88    & -23088.63112 & -23092.34870 & -23036.45668 & -23036.02407 \\
      \br
    \end{tabular}
  \end{indented}
\end{table}
%
%
\section{Relativistic Correction of Hartree Term}
\label{sec:app_hartree}
\par
In this section, the derivation of $ E_{\urm{Hrel}} $ (equation \eqref{eq:EH1}) is appended.
We define the contribution from the first and second terms of $ V'_2 $ as $ E_{\urm{H1}} $ and $ E_{\urm{H2}} $, respectively,
as $ E_{\urm{Hrel}} = E_{\urm{H1}} + E_{\urm{H2}} $;
\begin{align}
  & E_{\urm{H1}} \left[ \rho \right]
    \notag \\
  = & \, 
      - \frac{\pi}{2c^2}
      \sum_{j \ne k}^{\text{occ}}
      \iint
      \psi^*_j \left( \ve{r} \right) \,
      \psi^*_k \left( \ve{r}' \right) \,
      \delta \left( \ve{r} - \ve{r}' \right) \,
      \psi_j \left( \ve{r} \right) \,
      \psi_k \left( \ve{r}' \right) \,
      d \ve{r} \, d \ve{r}',
      \label{eq:Appendix_H1} \\
  & E_{\urm{H2}} \left[ \rho \right]
    \notag \\
  = & \, 
      -
      \frac{1}{4c^2}
      \sum_{j \ne k}^{\text{occ}}
      \iint
      \psi_j^* \left( \ve{r} \right) \,
      \psi_k^* \left( \ve{r}' \right) \,
      \overleftarrow{\nabla}
      \cdot
      \left[
      \frac{\left( \ve{r} - \ve{r}' \right) \left( \ve{r} - \ve{r}' \right)}{\left| \ve{r} - \ve{r}' \right|^3}
      +
      \frac{1}{\left| \ve{r} - \ve{r}' \right|}
      \right]
      \cdot
      \overrightarrow{\nabla}'
      \psi_j \left( \ve{r} \right) \,
      \psi_k \left( \ve{r}' \right) \,
      d \ve{r} \, d \ve{r}',
      \label{eq:Appendix_H2}
\end{align}
where the summation runs over the occupied states only.
The $ j = k $ contribution can be included in equations \eqref{eq:Appendix_H1} and \eqref{eq:Appendix_H2} 
since they are canceled by the exchange terms, 
and thus $ \sum_{j \ne k} $ can be replaced to $ \sum_{j, k} $.
Here, we derive the density functional forms of $ E_{\urm{H1}} $ and $ E_{\urm{H2}} $.
\par
The first term $ E_{\urm{H1}} $ is straightforwardly transformed as 
\begin{align}
  E_{\urm{H1}} \left[ \rho \right] 
  & =
    - \frac{\pi}{2c^2}
    \sum_{j, k}^{\text{occ}}
    \iint
    \psi^*_j \left( \ve{r} \right) \,
    \psi^*_k \left( \ve{r}' \right) \,
    \delta \left( \ve{r} - \ve{r}' \right) \,
    \psi_j \left( \ve{r} \right) \,
    \psi_k \left( \ve{r}' \right) \,
    d \ve{r} \, d \ve{r}' \notag \\
  & = 
    - \frac{\pi}{2c^2}
    \iint
    \rho \left( \ve{r} \right) \,
    \rho \left( \ve{r}' \right) \,
    \delta \left( \ve{r} - \ve{r}' \right) \,
    d \ve{r} \, d \ve{r}' \notag \\
  & = 
    - \frac{\pi}{2c^2}
    \int
    \left[
    \rho \left( \ve{r} \right) 
    \right]^2
    \, d \ve{r} .
\end{align}
\par
Next, relativistic correction $ E_{\urm{H2}} $ is derived with the assumption that the system has the time-reversal symmetry.
Here, the density is written with the single-particle Kohn--Sham orbital as 
\begin{equation}
  \rho \left( \ve{r} \right)
  = 
  \sum_j^{\text{occ}}
  \psi_j^* \left( \ve{r} \right) \,
  \psi_j \left( \ve{r} \right).
  \label{eq:dens_sp1}
\end{equation}
The time-reversal symmetry ensures that any complex conjugate of the occupied eigenstate
is also occupied eigenstate
and here its index is denoted as $ j^* $;
$ \psi_j^* \eqdef \psi_{j^*} $.
Thus, the component $ \sum_j^{\text{occ}} \left[ \nabla \psi_j^* \left( \ve{r} \right) \, \psi_j \left( \ve{r} \right) \right] $ can be written as
\begin{align}
  \sum_j^{\text{occ}} 
  \left[ \nabla \psi_j^* \left( \ve{r} \right) \, \psi_j \left( \ve{r} \right) \right]
  & =
    \frac{1}{2}
    \sum_j^{\text{occ}} 
    \left[ 
    \nabla \psi_j^* \left( \ve{r} \right) \, \psi_j \left( \ve{r} \right)
    +
    \nabla \psi_{j^*}^* \left( \ve{r} \right) \, \psi_{j^*} \left( \ve{r} \right)
    \right]
    \notag \\
  & =
    \frac{1}{2}
    \sum_j^{\text{occ}} 
    \left[ 
    \nabla \psi_j^* \left( \ve{r} \right) \, \psi_j \left( \ve{r} \right)
    +
    \nabla \psi_j \left( \ve{r} \right) \, \psi_j^* \left( \ve{r} \right)
    \right]
    \notag \\
  & = 
    \frac{1}{2}
    \nabla
    \sum_j^{\text{occ}} 
    \left|
    \psi_j \left( \ve{r} \right)
    \right|^2
    \notag \\
  & =
    \frac{1}{2}
    \nabla
    \rho \left( \ve{r} \right).
\end{align}
Equation~\eqref{eq:Appendix_H2}, hence, reads
\begin{align}
  & E_{\urm{H2}} \left[ \rho \right]
    \notag \\
  = & \, 
      -
      \frac{1}{4c^2}
      \sum_{j, k}^{\text{occ}}
      \iint
      \psi_j^* \left( \ve{r} \right) \,
      \psi_k^* \left( \ve{r}' \right) \,
      \overleftarrow{\nabla}
      \cdot
      \left[
      \frac{\left( \ve{r} - \ve{r}' \right) \left( \ve{r} - \ve{r}' \right)}{\left| \ve{r} - \ve{r}' \right|^3}
      +
      \frac{1}{\left| \ve{r} - \ve{r}' \right|}
      \right]
      \cdot
      \overrightarrow{\nabla}'
      \psi_j \left( \ve{r} \right) \,
      \psi_k \left( \ve{r}' \right) \,
      d \ve{r} \, d \ve{r}'
      \notag \\
  = & \,
      -
      \frac{1}{4c^2}
      \sum_{j, k}^{\text{occ}}
      \iint
      \psi_k^* \left( \ve{r}' \right) \,
      \frac{
      \left\{
      \left( \ve{r} - \ve{r}' \right)
      \cdot \nabla \psi_j^* \left( \ve{r} \right)
      \right\}
      \left\{
      \left( \ve{r} - \ve{r}' \right)
      \cdot \nabla' \psi_k \left( \ve{r}' \right)
      \right\}}
      {\left| \ve{r} - \ve{r}' \right|^3}
      \cdot
      \psi_j \left( \ve{r} \right) \,
      d \ve{r} \, d \ve{r}'
      \notag \\
  & \,
    -
    \frac{1}{4c^2}
    \sum_{j, k}^{\text{occ}}
    \iint
    \psi_k^* \left( \ve{r}' \right) \,
    \frac{
    \left\{
    \nabla \psi_j^* \left( \ve{r} \right)
    \right\}
    \cdot
    \left\{
    \nabla' \psi_k \left( \ve{r}' \right)
    \right\}}
    {\left| \ve{r} - \ve{r}' \right|}
    \cdot
    \psi_j \left( \ve{r} \right) \,
    d \ve{r} \, d \ve{r}'
    \notag \\
  = & \,
      -
      \frac{1}{16 c^2}
      \iint
      \left[
      \frac{
      \left\{
      \left( \ve{r} - \ve{r}' \right)
      \cdot \nabla \rho \left( \ve{r} \right)
      \right\}
      \left\{
      \left( \ve{r} - \ve{r}' \right)
      \cdot \nabla' \rho \left( \ve{r}' \right)
      \right\}}
      {\left| \ve{r} - \ve{r}' \right|^3}
      +
      \frac{
      \left\{
      \nabla \rho \left( \ve{r} \right)
      \right\}
      \cdot
      \left\{
      \nabla' \rho \left( \ve{r}' \right)
      \right\}}
      {\left| \ve{r} - \ve{r}' \right|}
      \right]
      d \ve{r} \, d \ve{r}'.
      \label{eq:Hartree_ret0}
\end{align}
\par
Here, in the atomic systems, the density $ \rho $ satisfies $ \rho \left( \ve{r} \right) \to 0 $ in $ r \to \infty $.
Under this assumption,
since
\begin{equation}
  \partial_j
  \frac{1}{\left| \ve{r} - \ve{r}' \right|}
  =
  - \frac{\ve{r}_j - \ve{r}'_j}{\left| \ve{r} - \ve{r}' \right|^3}
  \qquad
  \text{($ j = x, \, y, \, z $)},
  \label{eq:EH2_1}
\end{equation}
the second term of equation \eqref{eq:Hartree_ret0} reads
\begin{align}
  \iint
  \frac{
  \left\{
  \nabla \rho \left( \ve{r} \right)
  \right\}
  \cdot
  \left\{
  \nabla' \rho \left( \ve{r}' \right)
  \right\}}
  {\left| \ve{r} - \ve{r}' \right|}
  \, d \ve{r} \, d \ve{r}'
  & =
    -
    \iint
    \rho \left( \ve{r} \right)
    \nabla
    \frac{
    \nabla' \rho \left( \ve{r}' \right)}
    {\left| \ve{r} - \ve{r}' \right|}
    \, d \ve{r} \, d \ve{r}'
    \notag \\
  & =
    \iint
    \rho \left( \ve{r} \right)
    \frac{
    \left( \ve{r} - \ve{r}' \right)
    \cdot
    \left\{
    \nabla' \rho \left( \ve{r}' \right) \right\}}
    {\left| \ve{r} - \ve{r}' \right|^3}
    \, d \ve{r} \, d \ve{r}'.
    \label{eq:Hartree_ret2}
\end{align}
Since
\begin{equation}
  \partial_i
  \frac{\left( r_i - r'_i \right) \left( r_j - r'_j \right)}{\left| \ve{r} - \ve{r}' \right|^3}
  = 
  \frac{r_j - r'_j}{\left| \ve{r} - \ve{r}' \right|^3}
  +
  \frac{r_i - r'_i}{\left| \ve{r} - \ve{r}' \right|^3} \delta_{ij}
  -
  3 \frac{\left( r_i - r'_i \right)^2 \left( r_j - r'_j \right)}{\left| \ve{r} - \ve{r}' \right|^5}
  \label{eq:EH2_2}
\end{equation}
and
\begin{align}
  & 
    \sum_{i, \, j = 1}^3
    \partial_i
    \frac{\left( r_i - r'_i \right) \left( r_j - r'_j \right)}{\left| \ve{r} - \ve{r}' \right|^3}
    \partial'_j \rho \left( \ve{r}' \right)
    \notag \\
  = & \,
      \sum_{i, \, j = 1}^3
      \left(
      \frac{r_j - r'_j}{\left| \ve{r} - \ve{r}' \right|^3}
      +
      \frac{r_i - r'_i}{\left| \ve{r} - \ve{r}' \right|^3} \delta_{ij}
      -
      3 \frac{\left( r_i - r'_i \right)^2 \left( r_j - r'_j \right)}{\left| \ve{r} - \ve{r}' \right|^5}
      \right)
      \partial'_j \rho \left( \ve{r}' \right)
      \notag \\
  = & \,
      3
      \sum_{j = 1}^3 
      \frac{r_j - r'_j}{\left| \ve{r} - \ve{r}' \right|^3}
      \partial'_j \rho \left( \ve{r}' \right)
      +
      \sum_{i = 1}^3
      \frac{r_i - r'_i}{\left| \ve{r} - \ve{r}' \right|^3}
      \partial'_i \rho \left( \ve{r}' \right)
      -
      3
      \sum_{i, \, j = 1}^3
      \frac{\left( r_i - r'_i \right)^2 \left( r_j - r'_j \right)}{\left| \ve{r} - \ve{r}' \right|^5}
      \partial'_j \rho \left( \ve{r}' \right)
      \notag \\
  = & \,
      4
      \frac{\left( \ve{r} - \ve{r}' \right) \cdot \left\{ \nabla' \rho \left( \ve{r}' \right) \right\}}
      {\left| \ve{r} - \ve{r}' \right|^3}
      -
      3
      \sum_{i, \, j = 1}^3
      \frac{\left| \ve{r} - \ve{r}' \right|^2
      \left( \ve{r} - \ve{r}' \right) \cdot \left\{ \nabla' \rho \left( \ve{r}' \right) \right\}}
      {\left| \ve{r} - \ve{r}' \right|^5}
      \notag \\
  = & \,
      \frac{\left( \ve{r} - \ve{r}' \right) \cdot \left\{ \nabla' \rho \left( \ve{r}' \right) \right\}}
      {\left| \ve{r} - \ve{r}' \right|^3},
      \label{eq:EH2_3}
\end{align}
the first term of equation \eqref{eq:Hartree_ret0} reads
\begin{align}
  & 
    \iint
    \frac{
    \left\{
    \left( \ve{r} - \ve{r}' \right)
    \cdot \nabla \rho \left( \ve{r} \right)
    \right\}
    \left\{
    \left( \ve{r} - \ve{r}' \right)
    \cdot \nabla' \rho \left( \ve{r}' \right)
    \right\}}
    {\left| \ve{r} - \ve{r}' \right|^3}
    \, d \ve{r} \, d \ve{r}'
    \notag \\
  = & \,
      \iint
      \sum_{i, \, j = 1}^3
      \frac{
      \left\{
      \left( r_i - r'_i \right)
      \partial_i \rho \left( \ve{r} \right)
      \right\}
      \left\{
      \left( r_j - r'_j \right)
      \partial'_j \rho \left( \ve{r}' \right)
      \right\}}
      {\left| \ve{r} - \ve{r}' \right|^3}
      \, d \ve{r} \, d \ve{r}'
      \notag \\
  = & \,
      \iint
      \sum_{i, \, j = 1}^3
      \frac{
      \left( r_i - r'_i \right)
      \left( r_j - r'_j \right)
      \partial_i \rho \left( \ve{r} \right)
      \partial'_j \rho \left( \ve{r}' \right)}
      {\left| \ve{r} - \ve{r}' \right|^3}
      \, d \ve{r} \, d \ve{r}'
      \notag \\
  = & \,
      -
      \iint
      \sum_{i, \, j = 1}^3
      \rho \left( \ve{r} \right)
      \partial_i
      \frac{
      \left( r_i - r'_i \right)
      \left( r_j - r'_j \right)}
      {\left| \ve{r} - \ve{r}' \right|^3}
      \partial'_j \rho \left( \ve{r}' \right)
      \, d \ve{r} \, d \ve{r}'
      \notag \\
  = & \,
      -
      \iint
      \rho \left( \ve{r} \right)
      \frac{\left( \ve{r} - \ve{r}' \right) \cdot \left\{ \nabla' \rho \left( \ve{r}' \right) \right\}}
      {\left| \ve{r} - \ve{r}' \right|^3}
      \, d \ve{r} \, d \ve{r}'.
      \label{eq:Hartree_ret1}
\end{align}
Therefore, equation \eqref{eq:Hartree_ret0} reads
\begin{align}
  & 
    E_{\urm{H2}} \left[ \rho \right]
    \notag \\
  = & \,
      -
      \frac{1}{16 c^2}
      \iint
      \left[
      \frac{
      \left\{
      \left( \ve{r} - \ve{r}' \right)
      \cdot \nabla \rho \left( \ve{r} \right)
      \right\}
      \left\{
      \left( \ve{r} - \ve{r}' \right)
      \cdot \nabla' \rho \left( \ve{r}' \right)
      \right\}}
      {\left| \ve{r} - \ve{r}' \right|^3}
      +
      \frac{
      \left\{
      \nabla \rho \left( \ve{r} \right)
      \right\}
      \cdot
      \left\{
      \nabla' \rho \left( \ve{r}' \right)
      \right\}}
      {\left| \ve{r} - \ve{r}' \right|}
      \right]
      d \ve{r} \, d \ve{r}'
      \notag \\
  = & \,
      -
      \iint
      \rho \left( \ve{r} \right)
      \frac{\left( \ve{r} - \ve{r}' \right) \cdot \left\{ \nabla' \rho \left( \ve{r}' \right) \right\}}
      {\left| \ve{r} - \ve{r}' \right|^3}
      \, d \ve{r} \, d \ve{r}'
      +
      \iint
      \rho \left( \ve{r} \right)
      \frac{
      \left( \ve{r} - \ve{r}' \right)
      \cdot
      \left\{
      \nabla' \rho \left( \ve{r}' \right) \right\}}
      {\left| \ve{r} - \ve{r}' \right|^3}
      \, d \ve{r} \, d \ve{r}'
      \notag \\
  = & \,
      0.
\end{align}
\par
Finally,
\begin{equation}
  E_{\urm{Hrel}} \left[ \rho \right]
  = 
  E_{\urm{H1}} \left[ \rho \right] 
  =
  - \frac{\pi}{2c^2}
  \int
  \left[
    \rho \left( \ve{r} \right) 
  \right]^2
  \, d \ve{r} 
\end{equation}
is followed.
%
%
\providecommand{\newblock}{}

\end{document}